\documentclass[preprint]{revtex4}

\usepackage{graphicx}

\begin{document}

\title{The statistics of electric field fluctuations in liquid water}

\author{Bernhard Reischl, J\"urgen K\"ofinger, and Christoph Dellago \thanks{$^\ast$Corresponding author. Email: christoph.dellago@univie.ac.at}\\\vspace{6pt}
\vspace{6pt} {\em{Faculty of Physics and Center for Computational Materials Science, University of Vienna, Boltzmanngasse 5, 1090 Vienna, Austria}} }

\begin{abstract}
Electric field fluctuations play a major role in dissociation reactions in liquid water and determine its vibrational spectroscopic response. Here, we study the statistics of electric fields in liquid water using molecular dynamics computer simulations with a particular focus on the strong but rare fields that drive dissociation. Our simulations indicate that the important contributions to the electric field acting on OH bonds stem from water molecules less than 7~\AA{} away. Long-ranged contributions play a minor role. 
\end{abstract}

\bigskip

\maketitle

\section{Introduction}

The chemistry of water plays a crucial role in a wide range of processes in nature and technology ranging from enzymatic processes and the metabolism of living cells to acid-base reactions and proton transfer in hydrogen fuel cells. The theoretical study of these reactions is complicated by the fact that often water molecules act as reactants and products, as well as solvent. The archetypal example of such a complex reaction is autoionisation in liquid water. Here, a water molecule dissociates forming a pair of hydronium and hydroxide ions that then separate determining the pH of water. 

The autoionisation reaction in liquid water has been studied in experiments from which the reaction rate constants for dissociation and recombination have been determined \cite{Eigen,Natzle}. While these experiments provided valuable information on the kinetics of the reaction, their space and time resolution was not sufficient to clarify the atomistic details of the reaction. In principle, this information can be extracted from molecular dynamics simulations with forces calculated ab intio via density functional theory. Such simulations are, however, complicated by the widely disparate time scales present in the problem; while the typical vibration frequency of OH bonds is of the order of tens of femtoseconds, it takes an intact water molecule about 10 hours to dissociate on average. This time scale problem has been recently overcome \cite{DellagoScience} using a combination of Car-Parrinello molecular dynamics \cite{CarParrinello} and transition path sampling \cite{tps2,tps}. This approach, which does not require the definition of a reaction coordinate as in previous {\em ab initio} simulations of autoionisation \cite{TroutParr1,TroutParr2}, permits to concentrate on the reactive event and generate unbiased dissociation trajectories. Analysis of such trajectories yielded the following reaction mechanism. The dissociation of the water molecule is initiated by a rare but strong electric field fluctuation generated by the surrounding water molecules. The nascent hydronium and hydroxide ions then separate along a chain of hydrogen bonds via Grotthuss transfer events \cite{Agmon}. When the hydrogen bond chain still connecting the two ions breaks, rapid recombination is prevented and from this state the ions can move apart further and separate completely. 

While these transition path sampling simulations produced a detailed and physically appealing picture of the dissociation process, they did not identify the origin of the electric field fluctuation involved in the first stages of the dissociation process. In particular, no local hydrogen bonding pattern or ion coordination number could account for the presence of the bond-breaking electric field fluctuation. Also, it remained uncertain whether these electric fields, sufficiently strong to initiate the dissociation, stem from water molecules near the dissociating one or if long ranged effects are important. 

In this paper we study electric field fluctuations in liquid water using the TIP4P model \cite{Jorgensen} and investigate whether long-range electrostatics or specific local configurations are responsible for the strong but rare electric fields involved in the autoionisation process. In this particular model, water molecules are treated as rigid and the electrostatic interactions between molecules are modeled by point charges placed on the molecules. Since the model does not describe the cleavage and formation of covalent bonds, the dissociation event cannot be studied directly. Nevertheless, hydrogen bonds and long ranged electrostatic interactions are reproduced correctly such that information on the statistics of electric fields can be gleaned from our simulations. 

Recently, electric field fluctuations have also been considered in the context of the vibrational spectroscopy of liquid water \cite{Geissler,Geissler2,Hayashi,Skinner}. In such experiments, vibrational frequencies, sensitive to the local molecular environment, are used as probes for the dynamics of the hydrogen bond network percolating through the liquid. Using molecular dynamics simulations combined with a perturbative treatment of a simplified model of the OH oscillator, Geissler and collaborators \cite{Geissler} have shown that the frequency of the OH-stretch of a HOD molecule in liquid D$_2$O almost perfectly correlates with the electric field acting on the hydrogen site. An analysis of the microscopic molecular configurations revealed that the strength of the electric field, and hence the frequency of the OH-stretch, is mainly determined locally by the geometry of the HOD molecule with respect to the acceptor molecule; other molecules from the first solvation shell play only a minor role. At short times, vibrational decoherence is dominated by changes in this local geometry and only for longer times do collective density and polarization fluctuations, unrelated to specific molecular motions, determine the spectroscopic response. 

The remainder of this article is organised as follows. In Sec. \ref{sec:methods}, we described the simulation methodology and explain how we calculate and classify electric fields acting on OH bonds. Results are presented and discussed in Sec. \ref{sec:results} and conclusions are given in Sec. \ref{sec:conclusion}.

\section{Methods}
\label{sec:methods}

\subsection{Simulations}

	 As we are interested in electric field contributions from remote hydration shells, we need to study systems with at least $\approx 10^3$ water molecules. Currently, {\em ab initio} simulations with systems sizes of this magnitude are unfeasible, such that we have to resort to empirical water models. While such models do not describe bond breaking and forming, they capture effects due to long-ranged electrostatic interactions. In our simulations, we used the TIP4P interaction potential \cite{Jorgensen}, which reproduces the structure and dynamics of liquid water reasonably well, even compared to more recent potentials \cite{Guillot}.  The very good agreement between experiment and simulation obtained in Ref. \cite{Geissler2} for the spectroscopic response of HOD in liquid D$_2$O, which is closely related to the statistics of electric fields, confirms the suitability of empirical potentials to model electric field fluctuations in liquid water.
	 
	 The TIP4P model consists of four intermolecular interaction sites placed at fixed relative positions.  Two positive point charges are placed at the hydrogen atom centres and a compensating negative charge is placed on an extra point on the bisector of the HOH angle. In addition, a Lennard-Jones (LJ) interaction site is placed at the oxygen site. The total intermolecular interaction potential for two TIP4P water mole\-cu\-les $i$ and $j$ with respective charges $q_\alpha$ and $q_\beta$ and whose oxygen atoms are separated by the distance $R_{ij}$, can be written as
	\begin{equation}
 	v_{ij}=\sum_{\alpha}\sum_{\beta}\frac{1}{4\pi\varepsilon_0}\frac{q_{\alpha}q_{\beta}}{r_{i\alpha j\beta}}+\frac{A}{R_{ij}^{12}}-\frac{C}{R_{ij}^{6}},
	\end{equation}
 	where $r_{i\alpha j\beta}$ is the distance between the charge $\alpha$ on molecule $i$ and the charge $\beta$ on molecule $j$. The parameters of the TIP4P model are given in Tab.~\ref{TIP4Pparam}. 

	\begin{table}[!h]
	\begin{center}
 		\begin{tabular}{lll}
			\hline
			Symbol & Parameter & Value \\
			\hline\hline	
			$\alpha$ 		& HOH angle 		& $104.52^{\circ}$ \\
			$r_{\mathrm{OH}}$ 	& oxygen-hydrogen distance 	& 0.9572 \AA \\
			$r_{\mathrm{OM}}$ 	& oxygen-extra point distance 	& 0.15 \AA \\
			$q_{\mathrm{O}}$ 	& extra point charge 		& $-1.04$ e \\
			$q_{\mathrm{H}}$ 	& hydrogen charge		& +0.52 e\\
			$A$ 			& repulsive LJ parameter	& 0.6 \AA{}\textsuperscript{12} kcal/mol \\
			$C$ 			& attractive LJ parameter	& 610.0 \AA{}\textsuperscript{6} kcal/mol \\
			\hline
		\end{tabular}
		\caption{Parameters of the TIP4P potential.}
		\label{TIP4Pparam}
	\end{center}
	\end{table}

	All data presented in this paper were obtained from a 5 ns molecular dynamics simulations of 1000 water molecules performed at temperature 300 K in a cubic box with side length $L \approx 31.04$ \AA{} corresponding to a  density $\rho=1.0$ g/cm$^3$. The equations of motion were integrated using the velocity Verlet \cite{Swope} scheme with a time step of 0.5~fs and the RATTLE algorithm \cite{Rattle} was used to constrain bond lengths and keep the molecules rigid. The long ranged electrostatic interactions were calculated with Ewald summation \cite{Allen:compSim} with a real space cut-off of half the box size.

\subsection{Electric field calculation}
\label{sec:Efield}

	As we are interested in the electric fields involved in the dissociation of a water molecule, we study the electric fields at the centre of the molecule's OH bonds. Other researchers have studied electric field fluctuations at the center of charge of the molecule \cite{Hayashi} and at the hydrogen site \cite{Geissler2}. The latter choice is particularly relevant for the dynamics of OH-stretch, because the force acting on the OH-bond is a mass weighted sum of the forces on the oxygen atom and the hydrogen atom. Due to the smaller mass of the hydrogen atom, this sum is dominated by the hydrogen force, motivating the choice to study electric fields at the hydrogen site. We have, however, verified that the electric fields at the midpoint of the OH-bond and at the hydrogen position, while different in magnitude, are strongly correlated. In particular, our simulations indicate that unusually large electric fields occurring in these positions originate in the same molecular configurations.

	\begin{figure}[htb]
		\begin{center}   
			\includegraphics[angle=0, scale=0.3]{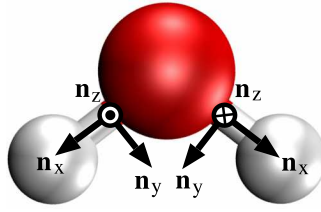}
			\caption{Cartesian coordinates located on the OH bonds. The origins lie on the centres of the OH bonds ($r_{\mathrm{OH}}$/2 $\approx $ 0.4786 \AA{}). The basis vectors $\mathbf{n}_{x}$ lie in direction of the OH bond, pointing towards the respective hydrogen atom. The basis vectors $\mathbf{n}_{y}$ lie also in the molecular plane, whereas the basis vectors $\mathbf{n}_{z}$ are perpendicular to the plane. (Visualisations of water molecules and simulation snapshots presented in this article were made with the VMD software package \cite{vmd}.)}
			\label{coordEbonds} 
		\end{center}
	\end{figure}

\begin{figure}[!h]
		\begin{center}   
			\includegraphics[angle=0, scale=0.30]{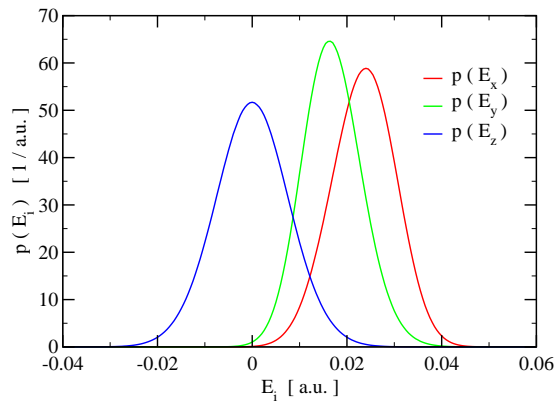}
			\caption{Distributions $p(E_{i})$ of the electric field components $E_x$, $E_y$, and $E_z$ at the centre of the OH bonds (see Fig.~\ref{coordEbonds}), 1 a.u.~$\approx$ 51.422 V/\AA{}.}
			\label{distEbonds} 
		\end{center}
	\end{figure}

	We place Cartesian coordinates in such a way, that the $x$-axes lie in direction of the OH bonds (see Fig.~\ref{coordEbonds}). A large positive value of the electric field in $x$-direction, $E_x$, will thus correspond to a force that pulls the hydrogen atom away from the oxygen atom. For a particular configuration, the electric field on the OH bond is calculated from the electrostatic forces acting on test charges located at the centre of the OH bonds, obtained from Ewald summation. The distributions of the electric field components are presented in Fig.~\ref{distEbonds}. Only $E_z$, the component perpendicular to the molecular plane, has a vanishing average, while both $E_x$ and $E_y$ have a positive average value. The field distributions are Gaussian to a high degree, indicating that the fields result from a sum of many contributions. The averages $\langle E_i \rangle$ and the standard deviations $\sigma(E_i)$ for the distribution functions of the three field components $E_x$, $E_y$, and $E_z$ are presented in Tab.~\ref{distEbondsfit}. It is interesting to note that the strong electric fields observed in the simulation generate forces of similar magnitude as the average forces acting during the dissociation event as computed by Trout and Parrinello using {\em ab initio} simulations \cite{TroutParr1,TroutParr2}. A direct comparison, however, is not possible since the rigid empirical potential used here does not permit the calculation of forces during the dissociation event, which is accompanied by a reorganisation of the charge distribution.  

\begin{figure}[!h]
		\begin{center}   
			\includegraphics[angle=0, scale=0.37]{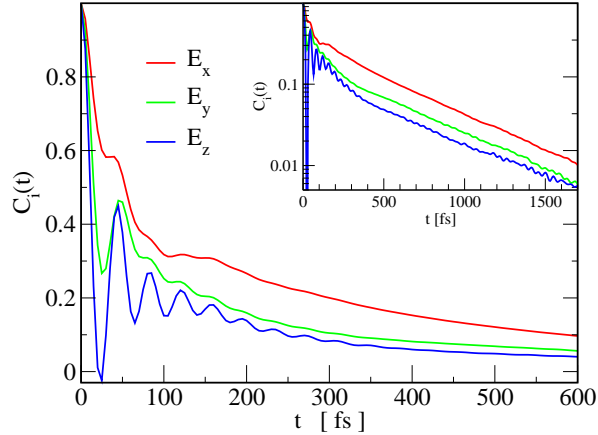}
			\caption{Normalised autocorrelation functions of the electric field components at the centre of the OH bonds (see Fig.~\ref{coordEbonds}). In the inset, the same autocorrelation functions are plotted with a logarithmic scale on the $y$-axis.}
			\label{EbondsACF} 
		\end{center}
	\end{figure}

The temporal persistence of electric field fluctuations can be inferred from the normalised time autocorrelation functions 
\begin{equation}
C_{i} (t)=\frac{ \langle E_i (0) E_i (t)\rangle - \langle E_i \rangle^2 } { \langle E_i^2 \rangle - \langle E_i \rangle {}^2 }
\end{equation}
depicted in Fig.~\ref{EbondsACF}. Here, the angular brackets denote equilibrium averages. The time autocorrelation functions of the three electric field components show distinct behaviour. In $z$-direction, perpendicular to the molecular plane, the autocorrelation function has pronounced oscillations with a period of about 40 fs, probably due to librational motions \cite{CowanNature}. The $x$-component shows a slightly longer decorrelation time than the other two components, but after $\approx250$ fs the autocorrelation functions of all three electric field components decay exponentially with a time constant $\tau\approx500$ fs. Our results are in good agreement with those of Mukamel and collaborators, who calculated electric field distributions and autocorrelation functions at the centre of charge of the molecule for different interaction potentials including TIP4P \cite{Hayashi}.

	\begin{table}[!h]
	 	\begin{center}
	 	 	\begin{tabular}{llll}
	 	 	 \hline
			 & $x$ & $y$ & $z$ \\
			\hline\hline
			$\langle E_i \rangle$ &  0.0237 & 0.0166 & 0.0000 \\
			$\sigma (E_i) $	& 6.77$\times10^{-3}$ &6.18$\times10^{-3}$ & 7.76$\times10^{-3}$ \\
			\hline
	 	 	\end{tabular}
			\caption{Gaussian fit parameters in atomic units for the distribution functions of the electric field components $E_x$, $E_y$, and $E_z$ on the OH bonds (see Fig.~\ref{distEbonds}).}
			\label{distEbondsfit}
	 	\end{center}
	\end{table}

\subsection{Shell Contributions}

In this section, we study how the observed electric field distributions are generated by the charges of the surrounding water molecules. To this end we introduce the shell contribution concept.

The total electric field $\mathbf{E}_{i \alpha}$ on test site $\alpha$ of molecule $i$ receives contributions from all charges $q_{j \beta}$ from molecules $j \neq i$ at distances $\mathbf{r}_{i \alpha j \beta}$,
	\begin{equation}
 	\mathbf{E}_{i \alpha} = \frac{1}{4\pi\varepsilon_0}\sum_{j\neq i} \sum_{\beta}  \frac{q_{j \beta}}{r_{i \alpha j \beta}^3} \mathbf{r}_{i \alpha j \beta}.
	\end{equation}
The  component of $\mathbf{E}_{i \alpha}$ along the OH bond is
	\begin{equation}
 	E \equiv E_x = \mathbf{E}_{i \alpha}\cdot \mathbf{n}^x_{i\alpha},
	\end{equation}
where $\mathbf{n}^x_{i\alpha}$ denotes the unit vector in the direction of the OH bond at test site $\alpha$ on molecule $i$. In the following we consider only the field component in $x$-direction and we omit the subscript $x$ for simplicity. 

\begin{figure}[!h]
		\begin{center}   
			\includegraphics[angle=0, scale=0.5]{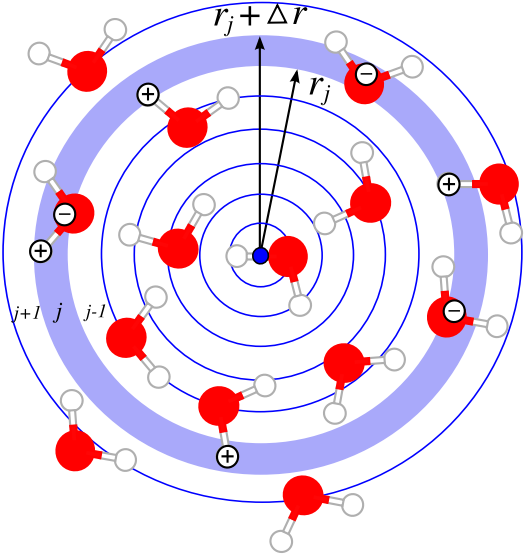}
			\caption{Shell contributions: in this two-dimensional illustration, the electric field on the centre of an OH bond receives contributions from positive (+) and negative (--) charges of the solvating water molecules. $E_{\mathrm{s}} (j)$ is the electric field originating from charges located in a spherical shell $j$. The shell $j$ is the volume (shaded region) between two concentric spheres with radii $r_j$ and $r_j+\Delta r$.}
			\label{shellcontribs} 
		\end{center}
	\end{figure}

	We now consider concentric spherical shells of width $\Delta r$ surrounding the test site on the OH bond (see Fig.~\ref{shellcontribs}). Every charge $q_i$ that is closer to the test site than the cut-off distance, will be located in one of these shells. For a given system configuration, some shells will contain one or more charges, whereas others will be empty. The electric field contribution $E_{\mathrm{s}} (j)$ of a certain shell $j$ to the $x$-component of the field at the test site is calculated as the sum over all electric field contributions $E_i$ from charges $q_i$ located in this shell,
\begin{equation}
E_{\mathrm{s}} (j)= \sum_{i=1}^{N_{\mathrm{c}}} E_i (\mathbf{r}_i) \Delta_j (\mathbf{r}_i).
\end{equation}
Here, $\mathbf{r}_i$ is the position of charge $q_i$ and $N_{\mathrm{c}}$ the total number of charges within the cut-off distance. The characteristic function $\Delta_j (\mathbf{r})$ is used to determine if a charge at position ${\bf r}$ is located within shell $j$, 
	\begin{equation}
	\Delta_j (\mathbf{r}) = \left \lbrace {1\atop 0}\quad {{\mathrm{if}\,\,\mathbf{r}\,\, \mathrm{is\,\,in\,\, shell\,\,}j}, \atop \mathrm{else.}} \right. 
	\end{equation}
In this way, the total electric field $E$ can be written as a sum over all shell contributions, 
	\begin{equation}
	E = \sum_{j=1}^{N_{\mathrm{s}}} E_{\mathrm{s}} (j),
	\end{equation}
where $N_{\mathrm{s}}$ the total number of shells.

We note that while the time evolution of the system is computed with electrostatic forces calculated via Ewald summation, direct summation of the electric fields is used for determining the shell contributions. For the system size considered here, the electrostatic forces obtained via Ewald and real space summation differ only slightly (see Tab.~\ref{fitshellewald}).

	To study the origin of the large electric field fluctuations, we calculate the distributions $P_j (\tilde{E_{\mathrm{s}}}|\tilde{E})$ of the electric field contribution $\tilde{E_{\mathrm{s}}}$ of shell $j$, given that the observed total electric field $E$ has a certain value $\tilde{E}$,
	\begin{equation}
	P_j (\tilde{E_{\mathrm{s}}}|\tilde{E}) = \frac{\langle\delta[\tilde{E_{\mathrm{s}}}-E_{\mathrm{s}} (j,r^N)]\,\delta[\tilde{E}-E(r^N)]\rangle}{\langle\delta[\tilde{E}-E(r^N)]\rangle}.
	\end{equation}
	Here, $r^N$ denotes a configuration of the system including the positions of all atoms. 
	From this distribution, we then evaluate the average contribution $\overline{E}_{\mathrm{s}} (j,\tilde{E})$ of shell $j$, for a given observed electric field $\tilde{E}$,
	\begin{eqnarray}
	\overline{E}_s (j|\tilde{E}) & = & \frac{\langle E_{\mathrm{s}} (j)\,\delta[\tilde{E}-E(r^N)]\rangle}{\langle\delta[\tilde{E}-E(r^N)]\rangle} \nonumber \\
	& = & \int \mathrm{d}\tilde{E}_{\mathrm{s}} P_j (\tilde{E_{\mathrm{s}}}| \tilde{E})\tilde{E_{\mathrm{s}}}.
	\end{eqnarray}
	In the computer simulation, rather than imposing a sharp value $\tilde{E}$ for the total electric field $E$, we determine the distributions for values of $E$ that are in a narrow interval $(\tilde{E}, \tilde{E}+\Delta E)$. 

	\begin{table}[!h]
	 	\begin{center}
	 	 	\begin{tabular}{llll}
	 	 	 \hline
			parameter & real space & Ewald sum & relative difference \\
			\hline\hline
			$\langle E \rangle$ &  0.0239 & 0.0237 & 1\% \\
			$\sigma (E) $	& 7.48$\times10^{-3}$ &6.77$\times10^{-3}$ & 10\% \\
			\hline
	 	 	\end{tabular}
			\caption{Comparison of the Gaussian fit parameters (in atomic units) for the distribution functions $p(E)$, obtained from shell contributions in real space and from the Ewald sum for a system size of $1000$ molecules.}
			\label{fitshellewald}
	 	\end{center}
	\end{table}

\section{Results and Discussion}
\label{sec:results}

\begin{figure}[!h]
		\begin{center}   
			\includegraphics[angle=0, scale=0.35]{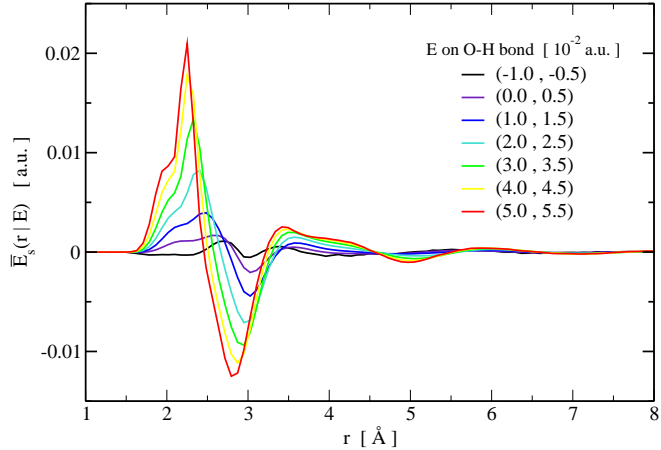}
			\caption{Average contributions $\overline{E}_{\mathrm{s}} (r | E)$ from charges in shells at distance $r$ to the total electric field $E$ at the OH bond centre, plotted as functions of $r$ for different ranges of the total electric field $E$ (values in parenthesis next to the line colour key).}
			\label{meanshellcontrib} 
		\end{center}
	\end{figure}

The average shell contributions for different values of the total field are depicted in Fig.~\ref{meanshellcontrib}. In this calculation we considered $N_{\mathrm{s}}=200$ concentric shells, with width $\Delta r \approx 0.075$ \AA. The largest shell radius corresponds to half the box size, $L/2 \approx 15.52$ \AA. Rather than using the shell index $j=1,\ldots,N_{\mathrm{s}}$, we denote the shells by their radial distance $r$ from the test site. The peaks in the curves of Fig.~\ref{meanshellcontrib} become more pronounced and are shifted to smaller radii for larger total fields.  For all considered values of the total electric field $E$, however, the non-zero contributions to the field come only from shells less than 7 \AA{} away from the test site on the OH bond centre. These correspond to the contributions of charges from molecules belonging to the first and second hydration shell of the water molecule hosting the test site. Hydration shells further away have no effective net contribution to the total field on the OH bond. This observation remains true even for large positive values of $E$, suggesting that rare electric fields involved in autoionisation in liquid water are a rather local effect.  	

\begin{figure}[!h]
		\begin{center}   
			\includegraphics[angle=0, scale=0.3]{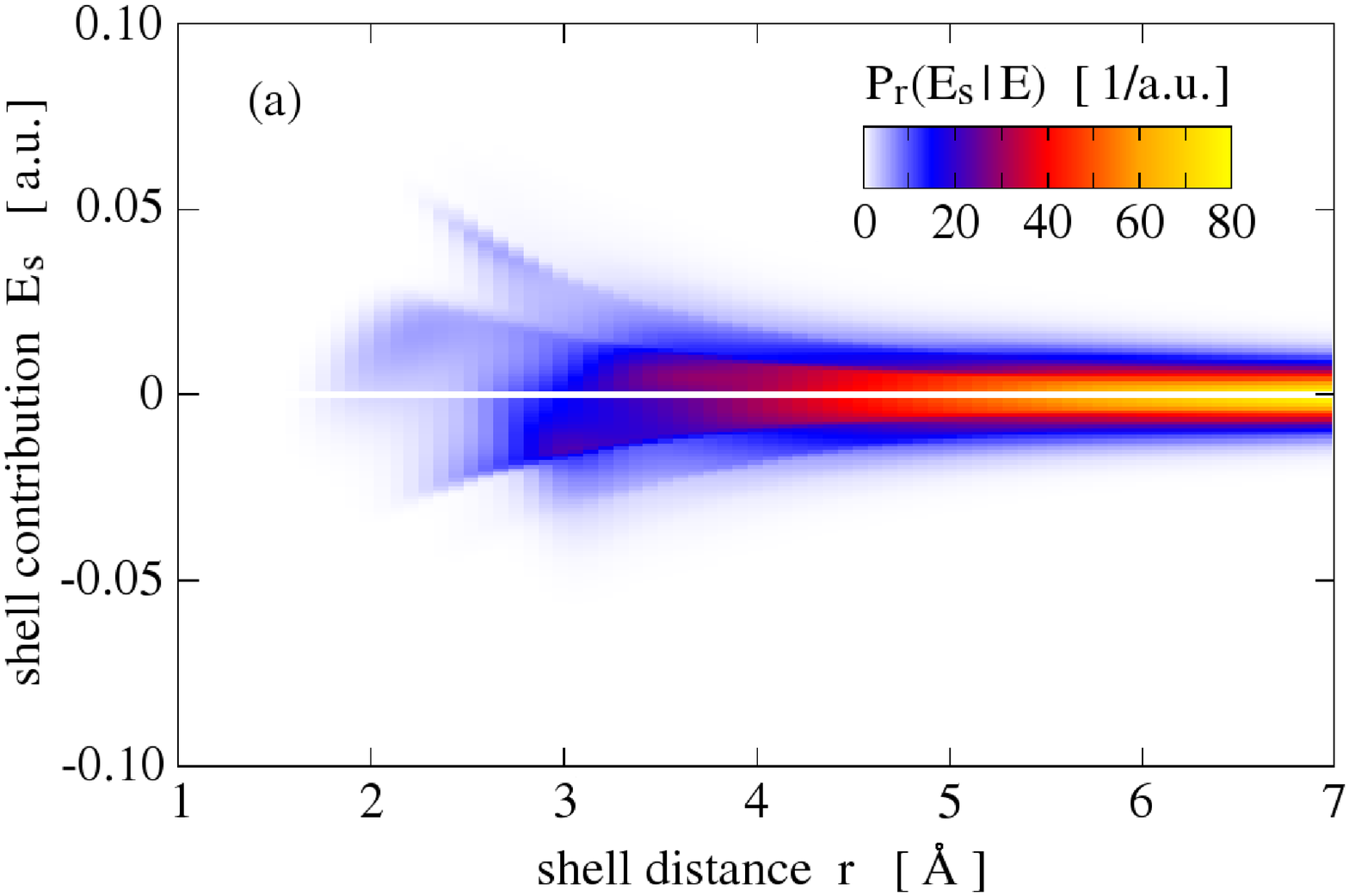}
			\includegraphics[angle=0, scale=0.3]{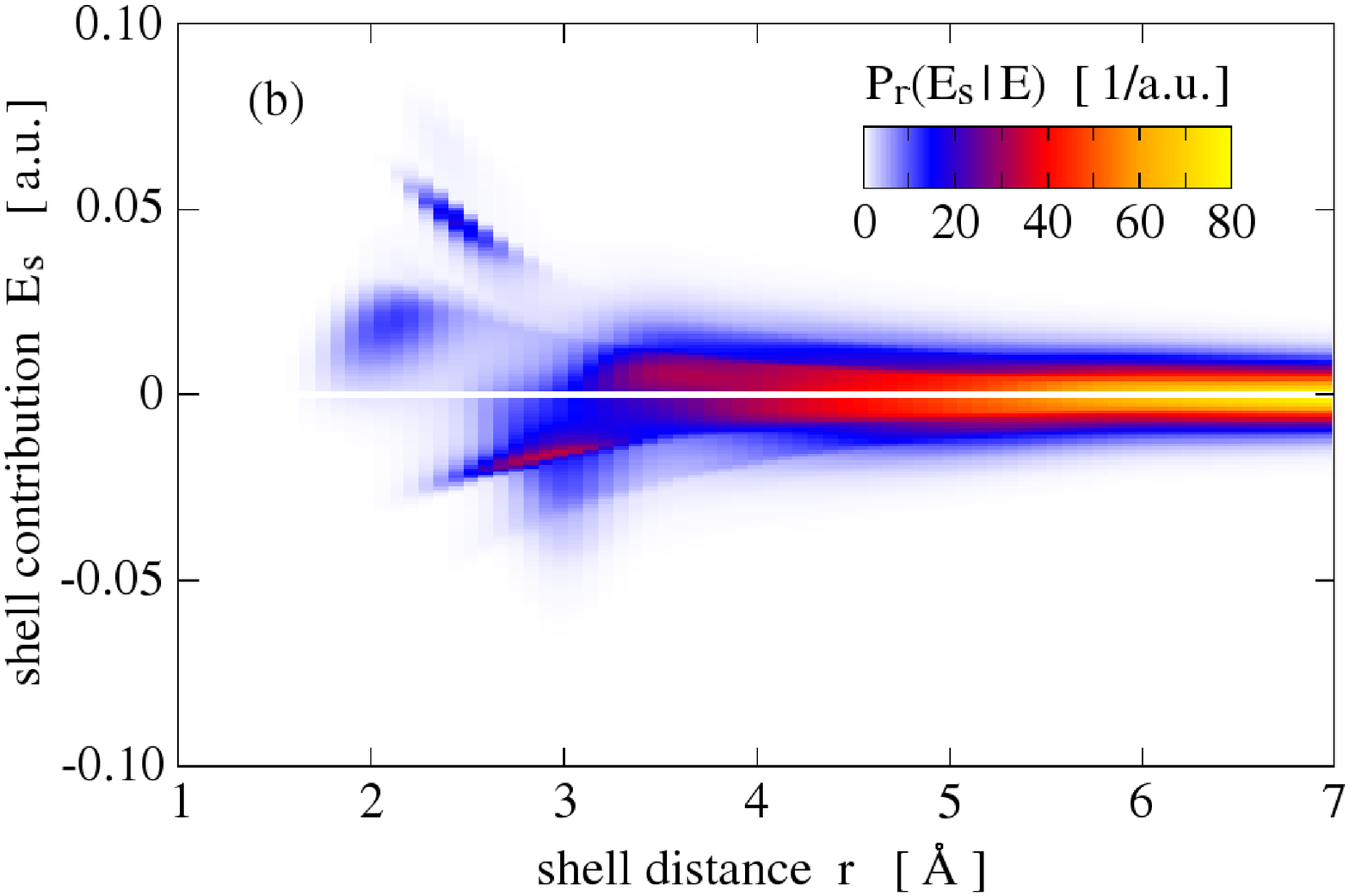}
			\includegraphics[angle=0, scale=0.3]{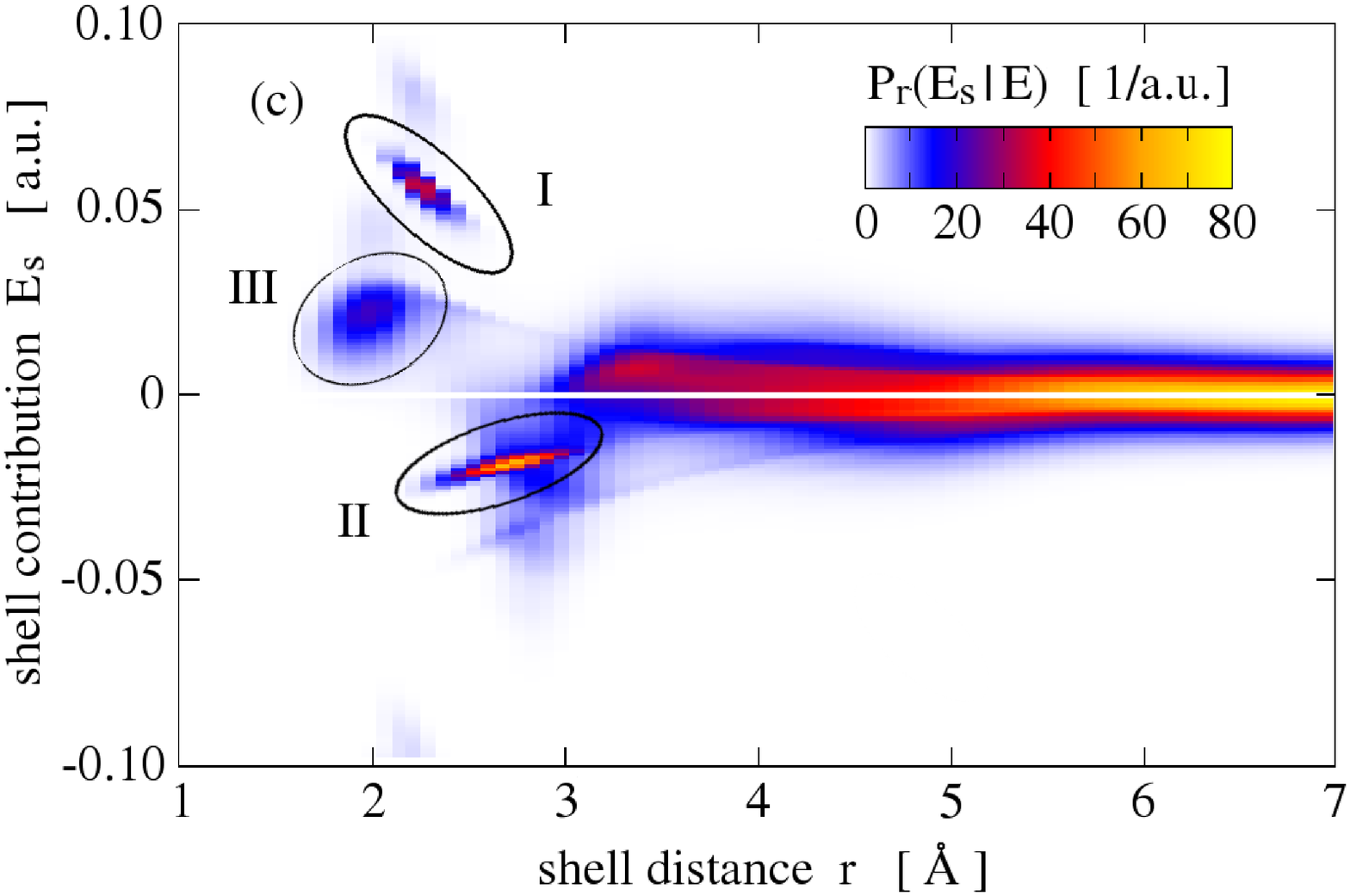}
			\caption{Shell contribution distributions $P_r (E_{\mathrm{s}} | E)$ as a function of shell distance $r$ and field contribution $E_{\mathrm{s}}$ for (a) weak total electric fields (0.000 -- 0.005 a.u.), (b) typical total electric fields (0.020 -- 0.025 a.u.) and (c) strong total electric fields (0.050 -- 0.055 a.u.). The peaks of $P_r (E_{\mathrm{s}} = 0 | E)$ for small shell distances $r$, caused by frequent absence of charges in shells with small volume, are omitted. The peaks in regions I -- III in graph (c) correspond to contributions from specific arrangements of water molecules as explained in detail in the main text.}
			\label{EFSCH} 
		\end{center}
	\end{figure}

The distributions of shell contributions, $P_j (\tilde{E_{\mathrm{s}}}|\tilde{E})$,  for various values of the total electric field are shown as colour coded maps in Fig.~\ref{EFSCH} for weak (a), average (b), and strong (c) electric fields. For very small observed values of the total field $E$ [Fig.~\ref{EFSCH}(a)], the average shell contributions are close to zero, even for small shell distances. Typical values of the total field [Fig.~\ref{EFSCH}(b)] are the result of positive contributions from shells at 1.7 to 2.7 \AA, and negative contributions from shells at 2.4 to 3.3 \AA. Shells further away yield only a small positive net contribution. Larger values of $E$ [Fig.~\ref{EFSCH}(c)] are generated in a similar way, but with larger and more strongly peaked positive and negative shell contributions. The contribution peaks are shifted towards smaller shell distances, yielding higher absolute values of $E$. For all values of $E$, shells at distances larger than 7 \AA{} show no significant contributions.

\subsection{Configuration analysis}

\begin{figure}[!h]
		\begin{center}   
			\includegraphics[angle=0, scale=0.5]{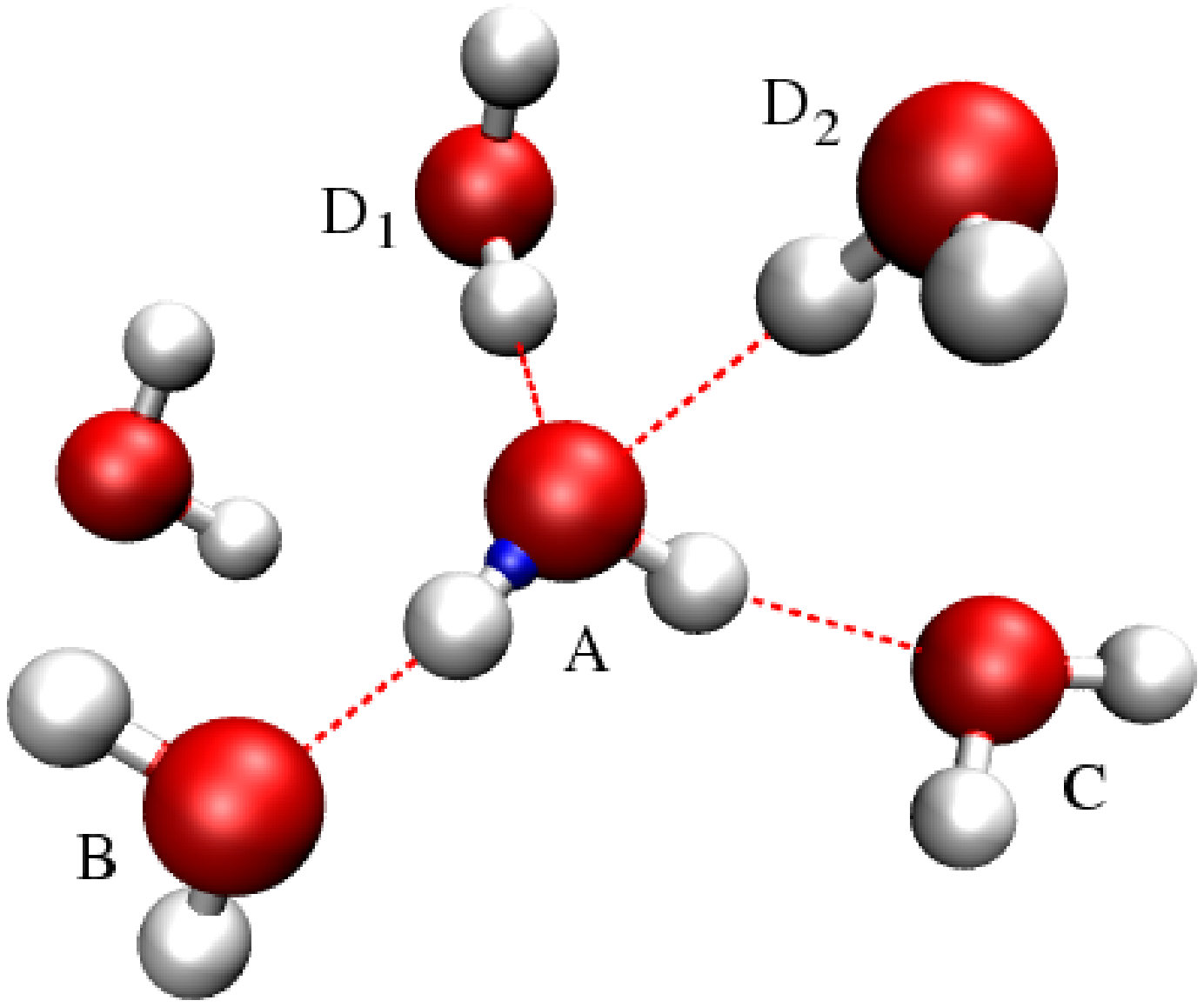}
			\caption{A typical configuration of water molecules generating a strong electric field, 0.050 a.u. $<E<$ 0.055 a.u., at the centre of the OH bond of the central molecule (dot on molecule A). Hydrogen bonds between water molecules are indicated by the dotted red lines. Molecule B accepts a hydrogen bond from molecule A involving the hydrogen next the test site. Molecule C accepts a hydrogen bond via the other hydrogen of molecule A. Molecules D$_1$ and D$_2$ donate hydrogen bonds to molecule A. }
			\label{config1} 
		\end{center}
	\end{figure}

	Sharp peaks in the distributions $P_r (E_{\mathrm{s}} | E)$, best visible in the case of large positive values of the total electric field $E$ [Fig.~\ref{EFSCH}(c)], indicate that typical arrangements exist where one or more charges at a certain distance produce a specific shell contribution $E_{\mathrm{s}}$. For small distances, the attribution of large electric fields to particular configurations of water molecules is geometrically possible. As the averaged contributions $\overline{E}_{\mathrm{s}} (r | E)$ rapidly decrease to values close to zero for shell distances $r>4$~\AA{} (Fig.~\ref{meanshellcontrib}), in our analysis we will only consider molecules with at least one charge closer than 4~{\AA} to the test site of the central molecule (molecule A in Fig.~\ref{config1}). 
	
	In most configurations that yield strong total electric fields in the range 0.050 a.u $< E <$ 0.055 a.u. the hydrogen atom next to the test site is involved in a hydrogen bond  to another water molecule (B in Fig.~\ref{config1}). The sharp peaks for shell radii between 2 and 3~\AA{} are caused by charges on this water molecule accepting the hydrogen bond. The peak at distance 2.1 -- 2.4~\AA{} and fields of 0.05 -- 0.07 a.u.~[region I in Fig.~\ref{EFSCH}(c)] originates from the negative charge $q_{\mathrm{O}}$. The other peak at distance 2.3 -- 3.0~\AA{} and fields of approximately $-0.02$ a.u.~[region II in Fig.~\ref{EFSCH}(c)] is due to the charge $q_{\mathrm{H}}$ of the hydrogen atom on the same molecule. Note that both of the curved ridged shapes created by the peaks of adjacent shells follow the $1/r^{2}$ dependence of the electric field. The predominance of molecule B in determining electric field fluctuations acting on the OH-bond was already pointed out in Ref. \cite{Geissler2}. As a consequence, vibrational spectroscopy of the OH-stretch is relatively insensitive to the motion of water molecules other than B and thus provides only a rather local probe of hydrogen bond dynamics. 

	Many configurations have one or two molecules (D${}_1$ and D${}_2$) donating hydrogen bonds to the oxygen atom of the molecule (A) hosting the test site. The diffuse peak at distance 1.8 -- 2.2~\AA{} and fields of 0.01 -- 0.03~a.u.~[region III in Fig.~\ref{EFSCH}(c)] can be attributed to the charges $q_{\mathrm{H}}$ on the hydrogen atoms of these molecules.

	The charges on a water molecule C accepting a hydrogen bond from the hydrogen atom opposite the test site show no significant contribution to the electric field $E$.

	The peaks in the distributions $P_r (E_{\mathrm{s}} | E)$ for strong fields, caused by the charges of the hydrogen bonded water molecule B adjacent to the test point, are essentially responsible for the positive and negative peaks at shell distances 2.2 -- 2.5~\AA{} and 2.5 --  3.0~\AA{}, respectively, in the averaged contributions $\overline{E}_{\mathrm{s}} (r | E)$ depicted in Fig.~\ref{meanshellcontrib}. The shoulder in the positive peak stems from the hydrogen charges of water molecules (D$_1$ and D$_2$) donating a hydrogen bond to A.

	\subsection{Local hydrogen bond patterns}
	
	Since the shell contributions can be traced back to the different types B, C, and D of molecules hydrogen bonded to the molecule A carrying the test site, we calculated the probabilities of the different local hydrogen bonding patterns for configurations yielding high values of the electric field and compared them to those configurations with average and low values of the electric field (see Tab.~\ref{configcontrib}). We denote the local hydrogen bond pattern of a molecule A by the number of molecules B, C and D bonded to it. For instance, '110' denotes a configuration with one hydrogen bond from A to B, one hydrogen bond from A to C, and no bonds from molecules D$_1$ and D$_2$ to A. Other hydrogen bond patterns are designated analogously. The geometric criterion chosen to detect hydrogen bonds in this analysis is that the oxygen-oxygen distance $r_{\mathrm{OO}}$ is less than 3.5 \AA{} and the angle $\phi$ between $\mathbf{r}_{\mathrm{OH}}$ of the donor and $\mathbf{r}_{\mathrm{OO}}$ is less than $30^{\circ}$. The specific figures in Tab.~\ref{configcontrib} depend on this choice quantitatively, but not qualitatively. Using these criteria, the average number of hydrogen bonds per water molecule is 3.52 as reported in earlier studies of TIP4P water \cite{JorgensenHBonds}.  

	Strong fields involve a higher number of hydrogen bonds. In this case, the molecule on average engages in 4.27 hydrogen bonds, with 1.79 bonds donated and 2.49 accepted. This reflects in patterns 112, 113, and 102 occurring most frequently. Configurations with less than three hydrogen bonds are not detected here. Average fields involve 3.54 hydrogen bonds, with 1.77 equally accepted and donated. Here, the most frequent configurations are 112 and 111. Except 112, patterns involving four or more hydrogen bonds are less frequent than in case of strong fields. In contrast, configurations with three hydrogen bonds or less become more frequent. Weak electric fields involve on average 2.38 hydrogen bonds, with 1.32 accepted and 1.06 donated. In this case, the configurations 111 and 011 are encountered most often. Patterns involving four or more hydrogen bonds occur only rarely, whereas configurations with two or less hydrogen bonds become frequent. An analysis carried out for equilibrium configurations of liquid water \cite{Skinner}, albeit with a different water model and a different hydrogen bond criterion, yielded similar frequencies of particular hydrogen bonding patterns as our analysis in the case of average electric field strengths. 
	
	It is interesting to note that for strong electric fields, in a large fraction of all configurations (more than 40\%) the water molecule with the test site accepts 3 hydrogen bonds. To a smaller degree, this over-coordination occurs also for typical fields, where about 5\% of all water molecules accept more than two hydrogen bonds. While the relative weight of such configurations depends on the particular hydrogen bond criterion used in the analysis, our results clearly indicate that anomalous hydrogen bonding patterns play an important role in the generation of large electric fields. Whether this behavior is a reflection of the real situation or is due to a deficiency of  the empirical force field used in our simulation is an interesting question that can be answered computationally only using more accurate potential energy surfaces such as those obtained from electronic structure calculations. 
	 
	Going from strong to weak fields, the average number of hydrogen bonds per molecule drops significantly from 4.27 to 2.38. In addition, the ratio of donated to accepted hydrogen bonds shifts: for strong fields, a molecule on average donates less hydrogen bonds than it accepts, whereas for weak fields the opposite can be observed. 
	
 	In summary, large electric fields on the test site strongly correlate with a higher than average number of hydrogen bonds and more bonds being accepted than donated by the water molecule carrying the test site.
	 
	\begin{table}[h]
	 	\begin{center}
	 	 	\begin{tabular}{ccccc}
	 	 	 \hline
			configuration & & \multicolumn{3}{c}{probability for}\\
			\cline{1-1} \cline{3-5}
			B C D & & weak field & average field & strong field\\
			\hline
			0 0 0 & & 0.013 & 0.000 & 0.000 \\
			0 0 1 & & 0.045 & 0.005 & 0.000 \\
			0 0 2 & & 0.012 & 0.010 & 0.001 \\
			\hline
			0 1 0 & & 0.074 & 0.000 & 0.000 \\
			0 1 1 & & 0.293	& 0.013	& 0.000 \\
			0 1 2 & & 0.110 & 0.056 & 0.000 \\
			\hline
			1 0 0 & & 0.016 & 0.003 & 0.000 \\
			1 0 1 & & 0.033 & 0.051 & 0.002 \\
			1 0 2 & & 0.010 & 0.075 & 0.130 \\
			1 0 3 & & 0.000 & 0.006 & 0.076 \\
			\hline
			1 1 0 & & 0.073 & 0.005 & 0.000 \\
			1 1 1 & & 0.222 & 0.210 & 0.003 \\
			1 1 2 & & 0.081 & 0.499 & 0.394 \\
			1 1 3 & & 0.004 & 0.051 & 0.367 \\
			1 1 4 & & 0.000 & 0.001 & 0.019 \\
			\hline
	 	 	\end{tabular}
			\caption{Probabilities of local hydrogen bond patterns, for configurations yielding weak (0.000 -- 0.005 a.u.), average (0.022 -- 0.026 a.u.), and strong (0.050 -- 0.055 a.u.) electric fields along the OH bond. The local configurations are defined by the number of different types of water molecules B, C and D hydrogen bonded to the molecule A on which the electric field is measured (see Fig.~\ref{config1}). Patterns with relative frequencies of less than 1\% are not shown.}
			\label{configcontrib}
	 	\end{center}
	\end{table}	

\section{Conclusion}
\label{sec:conclusion}

We have studied the distribution functions of the electric field, generated by solvent water molecules, on the centres of the OH bonds of a water molecule. In order to understand the origin of electric field fluctuations that are sufficiently strong to initiate the dissociation of a water molecule, we expressed the value of the electric field component along the bond as the sum of electric field contributions from charges in spherical shells around the centre of the OH bonds. By calculating the average shell contributions for different values of the observed field component, we were able to conclude that the relevant contributions come from shells at distances smaller than 7 \AA{}, corresponding to charges on water molecules of the first and second hydration shell, even for the largest values of the electric field observed in the simulation. This suggests that the mechanism leading to the autoionisation of water does not involve long-ranged rearrangements in the liquid. Indeed, particularly strong fields correlate with an increase in the number of hydrogen bonds to the molecule's neighbours and a higher number of hydrogen bonds being accepted than donated by the molecule on which the OH bond is considered.

Since the electric field acting on the water molecule has been shown to be almost uniquely related to the frequency of the OH-stretch \cite{Geissler2}, exceptionally strong fields may be accessible to experimental probes. Analysis of the respective molecular arrangements may then yield information on the likelihood of molecular patterns that drive chemical reactions through strong electric field fluctuations. 

All simulations discussed in this paper have been performed for rigid water molecules interacting via an empirical potential. The fields acting on a water molecule that is permitted to respond to the perturbation and actually undergo a dissociation are likely to be different from those reported here. This may be the reason why local hydrogen bond patterns are found to correlate with  strong fields in our simulations, but not in Ref. \cite{DellagoScience}. Further simulations with dissociable water molecules will be necessary to clarify this issue. Such studies require an {\em ab initio} calculation of forces and are becoming feasible on current computer systems for moderate system sizes. 

\section{Acknowledgements}

This work was supported by the Austrian Science Fund (FWF) under grant P20942-N16 and the Science College {\em Computational Materials Science} under grant W004, and by the University of Vienna through the University Focus Research Area {\em Materials Science} (project ``Multi-scale Simulations of Materials Properties and Processes in Materials'').


\begin{thebibliography}{20}

\bibitem{Eigen}
M. Eigen and L.D. Maeyer, Z. Electrochem. \textbf{59}, 986 (1955).

\bibitem{Natzle}
W.C. Natzle and C.B. Moore, J. Phys. Chem. \textbf{89}, 2605 (1985).

\bibitem{DellagoScience}
P.L. Geissler, C. Dellago, D. Chandler, J. Hutter, and M. Parrinello, Science \textbf{291},
  2121 (2001).

\bibitem{CarParrinello}
R. Car and M. Parrinello, Phys. Rev. Lett. \textbf{55}, 1985 (1985).

\bibitem{tps}
C. Dellago, P.G. Bolhuis and P.L. Geissler, Adv. Chem. Phys. \textbf{123}, 1
  (2002).

\bibitem{tps2}
C. Dellago, P.G. Bolhuis, F.S. Csajka and D. Chandler, J. Chem. Phys.
  \textbf{108}, 1964 (1998).

\bibitem{TroutParr1}
B. Trout and M. Parrinello, J. Phys. Chem. B \textbf{103}, 7340 (1999).

\bibitem{TroutParr2}
B. Trout and M. Parrinello, Chem. Phys. Lett. \textbf{288}, 343 (1999).

\bibitem{Agmon}
N. Agmon, Chem. Phys. Lett. \textbf{244}, 456 (1955).

\bibitem{Jorgensen}
W.L. Jorgensen, J. Chandrasekhar, J.D. Madura, R. W. Impey, and M. L. Klein, J. Chem. Phys.
  \textbf{79}, 926 (1983).

\bibitem{Hayashi}
T. Hayashi, T. la~Cour~Jansen, W. Zhuang and S. Mukamel, J. Phys. Chem. A
  \textbf{109}, 64 (2005).

\bibitem{Geissler}
C.J. Fecko, J.D. Eaves, J.J. Loparo, A. Tokmakoff, and P. L. Geissler, Science \textbf{301}, 1698
  (2003).

\bibitem{Geissler2}
J.D. Eaves, A. Tokmakoff, and P.L. Geissler
J. Chem Phys. A  {\bf 109}, 9424 (2005).

\bibitem{Skinner}
B. Auer, R. Kumar, J. R. Schmidt, and J. L Skinner,
Proc. Natl. Acad. Sci. USA \textbf{104}, 14215 (2007).

\bibitem{Guillot}
B. Guillot, J. Mol. Liq. \textbf{101}, 219 (2002).

\bibitem{Swope}
W.C. Swope, H.C. Andersen, P.H. Berens and K.R. Wilson, J. Chem. Phys.
  \textbf{76}, 637 (1982).

\bibitem{Rattle}
H.C. Andersen, J. Comp. Phys. \textbf{52}, 24 (1983).

\bibitem{Allen:compSim}
M.P. Allen and D.J. Tildesley, \emph{Computer Simulation of Liquids},
  $1^{\mathrm{st}}$ ed.   (Oxford University Press, Oxford, 2001).

\bibitem{vmd}
W. Humphrey, A. Dalke and K. Schulten, J. Mol. Graph. \textbf{14}, 33 (1996).

\bibitem{CowanNature}
M.L. Cowan, B.D. Bruner, N. Huse, J. R. Dwyer, B. Chugh, E. T. J. Nibbering, T. Elsaesser, and R. J. D. Miller , Nature \textbf{434}, 199 (2005).

\bibitem{JorgensenHBonds}
W.L. Jorgensen and J.D. Madura, Mol. Phys. \textbf{56}, 1381 (1985).

\end{thebibliography}
\end{document}